\pdfoutput=1
\documentclass[aps,prl,superscriptaddress,twocolumn]{revtex4-2}
\usepackage{float}
\usepackage{graphicx}
\usepackage{xcolor}
\usepackage{comment}

\begin{document}

\title{Electronic-mediated nuclear stopping power in proton irradiated water ice}

\author{Daniel Mu\~noz-Santiburcio}
\email[E-mail: ]{daniel.munozsan@upm.es}
\affiliation{CIC nanoGUNE BRTA, Tolosa Hiribidea 76, 20018 San Sebasti\'an, Spain}
\affiliation{Instituto de Fusi\'on Nuclear ``Guillermo Velarde'', 
Universidad Polit\'ecnica de Madrid, C/ Jos\'e Guti\'errez Abascal 2, 28006 Madrid, Spain}

\author{Jorge Kohanoff}
\affiliation{Instituto de Fusi\'on Nuclear ``Guillermo Velarde'',
Universidad Polit\'ecnica de Madrid, C/ Jos\'e Guti\'errez Abascal 2, 28006 Madrid, Spain}

\author{Emilio Artacho}
\affiliation{CIC nanoGUNE BRTA, Tolosa Hiribidea 76, 20018 San Sebasti\'an, Spain}
\affiliation{Theory of Condensed Matter, Cavendish Laboratory,
University of Cambridge, J. J. Thomson Ave, Cambridge CB3 0HE, United Kingdom}
\affiliation{Donostia International Physics Center (DIPC), Tolosa Hiribidea 76, 20018 San Sebasti\'an, Spain}
\affiliation{Ikerbasque, Basque Foundation for Science, 48011 Bilbao, Spain}

\date{\today}

\begin{abstract}
Traditionally, it has been assumed that the stopping of a swift ion travelling through matter can be understood in
terms of two 
essentially independent
components, i.e. electronic \textit{vs.} nuclear.
Performing 
extensive Ehrenfest MD simulations
of the process of proton irradiation of water ice
that accurately describe not only the non-adiabatic dynamics of the electrons but also of the nuclei, 
we have found 
a stopping mechanism involving the interplay of the  electronic and nuclear subsystems.
This effect, which consists in a kinetic energy transfer from the projectile to the target nuclei 
thanks to the perturbations of the electronic density caused by the irradiation, 
is fundamentally different
from the 
atomic displacements and collision cascades characteristic of nuclear stopping.
Moreover, 
it shows a marked isotopic effect depending on the composition of the target,
being relevant mostly for light water as opposed to heavy water.
This result is consistent with long-standing experimental results 
which remained unexplained so far.
\end{abstract}

%\keywords{}

\maketitle

The interaction of a swift ion with matter has been usually rationalized 
using
the concept of `stopping power', which consists in expressing the 
capacity of a target material to stop a given ion travelling through it 
as the energy lost by the projectile per unit of travelled length.
It has been usually assumed that this has two components,
being the `electronic' and `nuclear stopping power'~\cite{ziegler_book2008}. 
The former consists of the energy absorbed by the electronic subsystem of the target,
which becomes excited as a consequence of the projectile' passing,
while the latter consists of the energy gained by the target' nuclei
as a consequence of the direct impact of the projectile with them.
Both these quantities show a strong dependence on the velocity of the projectile,
in such a way that the velocity regimes where each of them has a significant 
value are basically non-overlapping. 
While it has been usually acknowledged that there is a certain domain of
projectile velocities where both effects coexist, 
it was assumed that this happened 
where both components are quite small,
and therefore this coincidence was not regarded as important for most purposes.
Especially for computer simulations aimed at determining the electronic stopping power
of any material by first-principles methods, the motion of the target' nuclei
and even the change of the velocity of the projectile are 
almost always neglected not only for practical reasons 
(i.e. alleviating the computational cost of the simulation
by skipping the calculation of the atomic forces), 
but also because 
nuclear stopping is assumed to be negligible in the domain of projectile velocities that are
relevant for the electronic stopping power.
In this work,
we show that such approximation neglects a channel for the projectile-target energy transfer
that has remained hitherto unexplored.

\textit{Methods.--}~
We performed Real-Time Time-Dependent DFT simulations within the Ehrenfest MD (EMD) formalism,
explicitly considering the forces on all atoms during the irradiation process, 
employing the CP2K code~\cite{cp2k_jcp2020} 
and using the Perdew-Burke-Ernzerhof (PBE)~\cite{Perdew1996+Erratum} exchange-correlation functional. 
All the simulations were carried out at the all-electron level 
with the \texttt{6-311++G2d2p} basis set~\cite{krishnan1980self,frisch1984self},
using the GAPW method~\cite{lippert_theorchemacc1999,krack_pccp2000}
with a cutoff of 350~Ry for the planewave expansion of the soft part of the density.
The model system consists of a Ih ice slab with 144 H$_2$O molecules in the unit cell
constructed from the corresponding bulk system with the experimental density at $T=145$~K
($\rho =931.19$~kg/m$^3$, with $a = 15.59$, $b=13.50$ and $c=21.98$~\AA~in
 the bulk~\cite{petrenko_book1999}, extended to $c=33.0$~\AA~in the slab), 
with a proton projectile initially placed in the vacuum region above the slab and with an initial 
velocity perpendicular to the slab.
We simulated 19 different proton trajectories for each one of 11 different values of the projectile' velocity,
from 0.2 to 8.0 a.u.
The timesteps for the EMD simulations were chosen to obtain a projectile displacement of
$\approx 0.005$~\AA~per EMD step in each case.
Further details will be published elsewhere, 
including a deeper description of the methods and further analyses of the results.

\begin{figure}[htb!]
        \includegraphics[width=.5\textwidth]{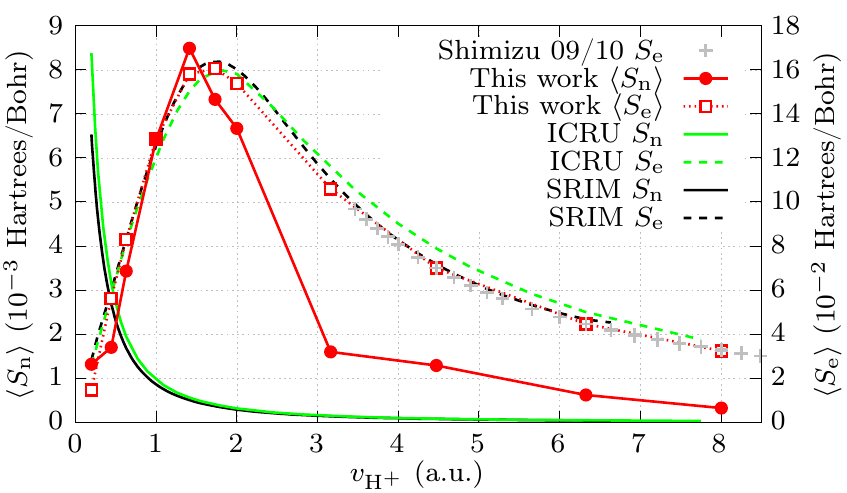}
        \caption{\label{fignucstop} Average nuclear and electronic stopping power 
 of protons in water ice computed in this work, 
compared to that obtained from SRIM 2013~\cite{ziegler_nimphysresb2010,srim2013}
(using a 0.94 compound correction) and ICRU 2014~\cite{pstar_icru2014},
in addition to the $S_\mathrm{e}$ experimental results of Shimizu \textit{et al}~\cite{shimizu_nimphysresb2009,shimizu_vacuum2010}. 
All values are scaled to a water density of 998 kg/m$^3$ for an easier comparison.}
\end{figure}

\textit{Stopping power: electronic vs. nuclear.--}~
The electronic, nuclear and total stopping power are readily obtained from each simulation as
$S_\mathrm{e} = \Delta E_\mathrm{KS}/d$,
$S_\mathrm{n} = \Delta E_\mathrm{kin}^\mathrm{target}/d$, and
$S_\mathrm{t} = - \Delta E_\mathrm{kin}^\mathrm{H^+}/d$,
being respectively the differences between the initial and final EMD steps of the 
Kohn-Sham energy and of the kinetic energies of the target nuclei and the projectile, 
in all cases divided by the 
distance travelled by the projectile within the target.
These definitions imply that in the following we will employ the terms `electronic' and `nuclear'
stopping as mere quantifyiers of the energy transferred to the electronic \textit{vs.} nuclear subsystems,
with independence of the underlying mechanism of such transfer.

In all the simulations, the relation $S_\mathrm{t} = S_\mathrm{e} + S_\mathrm{n}$ holds,
with an excellent energy conservation along the irradiation process.
As shown in Fig.~\ref{fignucstop},
the average electronic stopping power agrees remarkably well with that from
SRIM 2013~\cite{ziegler_nimphysresb2010,srim2013}
and with the tabulated data in ICRU~Report~90~\cite{pstar_icru2014},
with a perfect agreement 
with the experimental results of Shimizu~\textit{et al.}~\cite{shimizu_nimphysresb2009,shimizu_vacuum2010} at high velocities.
On the other hand, the average nuclear stopping power 
differs greatly from the
SRIM and ICRU data, 
occurring the greatest values in the same regime of projectile velocities as the electronic stopping maximum.
The small $\langle S_\mathrm{n} \rangle$ obtained by us in the low velocity limit 
is surely a consequence of the small size- and timescales in our simulations, 
which do not allow to observe the proper collision cascades which are characteristic in the regular nuclear stopping processes.
On the other hand, the strikingly high $\langle S_\mathrm{n} \rangle$ values in
the $v_\mathrm{H^+}$ range from 1 to 2~a.u. 
require a more complex explanation.

\begin{figure*}[htb!]
        \includegraphics[width=1.\textwidth]{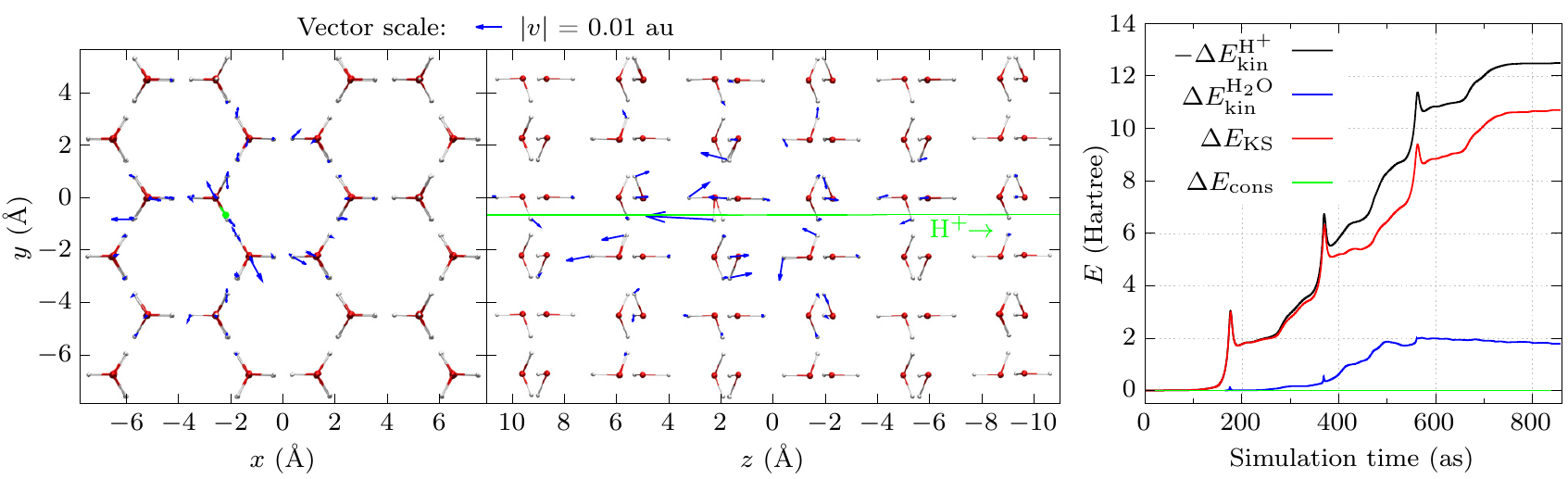}
        \caption{\label{figvel} Left/center: Atomic positions and velocities at the end of the simulation with the highest $S_\mathrm{n}$ in this work,
obtained with a 75~keV proton.
        The scale of the vectors indicating the velocities is shown above;
        vectors are only drawn for atoms with velocities greater than 0.001 a.u.
        The initial/final positions of the projectile are shown as a green circle in the $xy$ projection
        and its trajectory is shown as a green line in the $yz$ projection; note that the projectile travels in the $-z$ direction.
        Right: Change of the different energy components during the same simulation with respect to their initial values.}
\end{figure*}

The origin of such unexpected result can be found after careful inspection of all the individual EMD trajectories.
As we will expose in detail in further publications,
in few cases at the lowest projectile velocity,
the dynamics of the system is consistent with a `classical' mechanism 
where the target nuclei which gain more kinetic energy are those closer to the path of the projectile,
existing also a certain deflection of the projectile which supports the usual
picture of nuclear stopping in terms of elastic collisions.
However, as $v_\mathrm{H^+}$ increases, the behavior of the system is no longer fully consistent with that picture.
On one hand, the closest nuclei to the projectile path (with impact parameters of about $0.15$~\AA~or even less)
acquire velocities that are not consistent with a binary collision mechanism (Fig.~\ref{figvel}). 
Moreover, we find several nuclei quite far from the projectile path 
(belonging to molecules that are first or even second neighbors of the water molecules that are directly affected by the projectile) 
that gain a significant kinetic energy. In all the cases, the target nuclei that gain such kinetic energy are the hydrogen ones, 
being the kinetic energy gained by the oxygen nuclei negligible.

\begin{figure*}[htp!]
        \includegraphics[width=.6\textwidth]{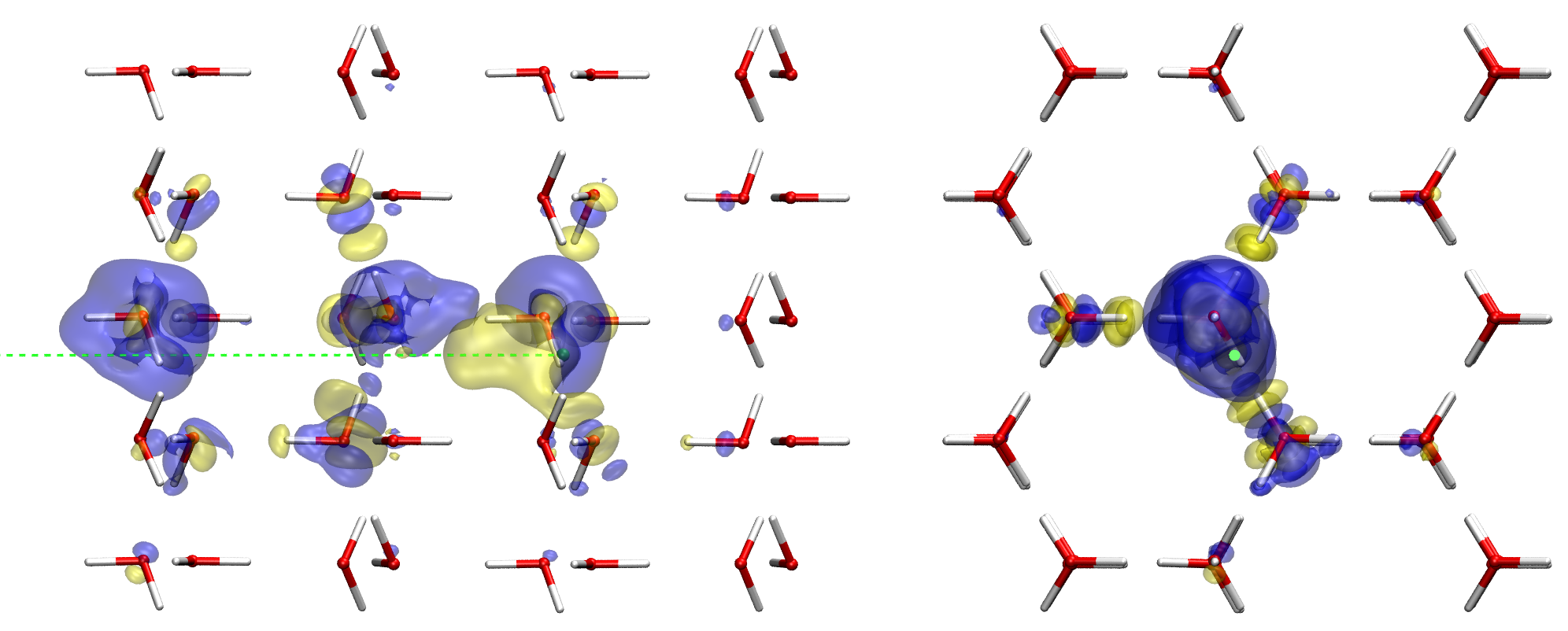}
        \includegraphics[width=.39\textwidth]{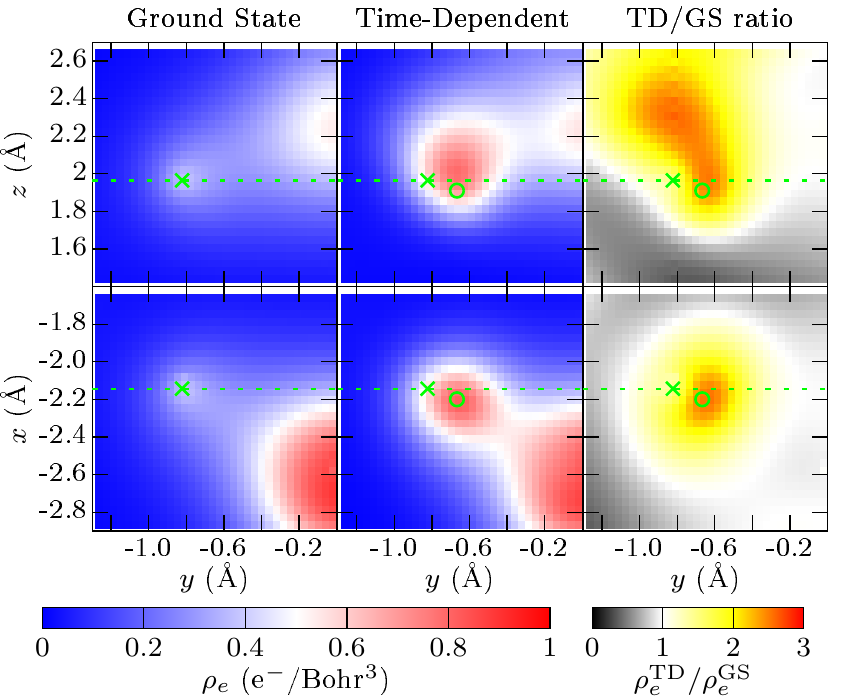}
        \caption{\label{figdens} Analysis of the electronic density for the simulation with the highest $S_\mathrm{n}$ at an instant where the projectile (green ball) has just passed close
        to the target atom with the highest kinetic energy increase. Left/center panels: side/top views of the system, with the blue/yellow isosurfaces (at an isovalue of $\pm 0.005$~e$^-$/Bohr$^3$)
        representing respectively the regions of density depletion/accumulation with respect the ground state, i.e. $\Delta \rho = \rho^\mathrm{TD} - \rho^\mathrm{GS}$, 
        with  $\Delta \rho < 0$ in blue and  $\Delta \rho > 0$ in yellow.
        Right panels: electronic density at clipping planes passing through the said target atom (green ``$\times$'' symbol) parallel to the $xy$/$yz$ directions,
        computed at the ground state \textit{vs.} that obtained during the TD-DFT irradiation simulation, showing also the ratio of the non-adiabatic to the ground state density.
        The position of the projectile is indicated as a green circle (note that the projectile is slightly offset from the clipping plane)}
\end{figure*}

\textit{Electronic density perturbations and nuclear stopping.--}~
The fact that the highest $\langle S_\mathrm{n} \rangle$ and $\langle S_\mathrm{e} \rangle$ values are both obtained
at the same $v_\mathrm{H^+}$ range
suggests that this unexpected energy transfer to the nuclei may be related 
to the perturbation of the target electronic subsystem during the irradiation process.
The analysis of 
the individual simulations 
confirms
a certain correlation between $S_\mathrm{n}$ and $S_\mathrm{e}$,
though the $S_\mathrm{n}/S_\mathrm{e}$ ratio varies significantly among the different trajectories, 
being 
as high as 15--20\% in some of them.
In particular,
$S_\mathrm{n}$ is higher in the cases where the projectile crosses, 
or passes very close to, the O--H bonds in the target. 
Inspection of the electronic density of the system during the irradiation process
for such cases reveals that the passing of the projectile causes 
very severe changes in the density distribution with respect to the ground state (Fig.~\ref{figdens}),
which can be as high as a threefold increase of the density in the close vicinity of the closest target atom.

Such perturbations of the electronic density of the water molecules 
along the projectile path
induce further density changes
that are propagated to the neighboring waters along the H-bond network, as visualized in Fig.~\ref{figdens}.
Analysis of the forces in such regions reveals that several H and O atoms therein experience extremely high forces,
though only the hydrogen atoms eventually acquire meaningful kinetic energies
(note that similar forces on H/O atoms imprint an acceleration on H that is 16-fold that on O,
which together with the short timescale of the process implies that only hydrogen atoms
can reach a significant velocity).
We note in passing that this effect does not involve the oxygen' core electrons in any way,
and actually 
thorough electronic structure analyses that will be unfolded elsewhere
show that the role of oxygen' core electrons in the proton stopping process is
much smaller than previously thought~\cite{yao_prl2019}.

\begin{figure}[htb!]
        \includegraphics[width=.5\textwidth]{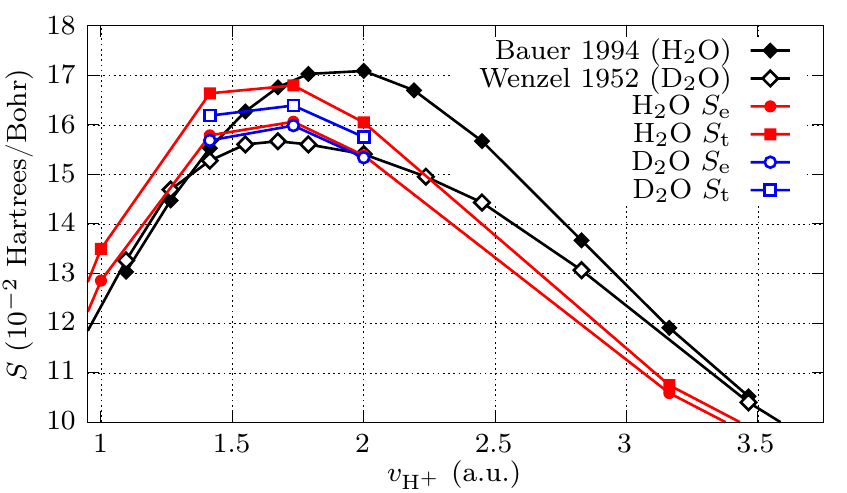}
        \caption{\label{figtotelstop} Electronic stopping power experimentally reported for H$_2$O~\cite{bauer_nimphysresb1994} 
        and D$_2$O~\cite{wenzel_pr1952} (respectively filled and empty diamonds)
        together with the average electronic and total stopping power (resp. circles/squares) found in this work from first-principles simulations
        for light and heavy ice (resp. red filled \textit{vs.} blue empty symbols).
        All values are scaled to correspond to a light water density of 998 kg/m$^3$ for an easier comparison.}
\end{figure}

\textit{Isotopic effect in electronic-mediated nuclear stopping.--}~
The findings above disclosed strongly suggest the possibility of an isotopic effect
for the stopping of protons in light vs.~heavy water, as the mass of the target nuclei
becomes relevant for this electronic-mediated nuclear stopping to take place.
Motivated by this, we performed additional irradiation simulations of protons in heavy ice,
employing the same simulation settings previously used for light ice,
for the three projectile velocities that showed the highest total stopping in light ice
($v_\mathrm{H^+} = 1.41$, 1.72 and 2.00~a.u.).
The results, in Fig.~\ref{figtotelstop}, show that the total stopping is indeed smaller in deuterated ice,
while the electronic stopping is practically the same.
Actually, long-standing results in the literature~\cite{wenzel_pr1952,bauer_nimphysresb1994} confirm the existence of an 
isotopic effect in the irradation of water by protons,
which has remained so far unexplored to the best of our knowledge.
Indeed, the precise isotopic composition of the target is usually ignored
when analyzing or comparing the electronic stopping power from different experiments or simulations,
due to the underlying assumption that $S_\mathrm{e}$ is solely determined by the response of the electronic subsystem of the target
and thus independent of the nuclear masses.

In Fig.~\ref{figtotelstop} we plot the electronic stopping power for proton projectiles
experimentally determined by Wenzel and Whaling using heavy water~\cite{wenzel_pr1952}
and the one by Bauer \textit{et al.} using light water~\cite{bauer_nimphysresb1994},
together with our simulation results.
Remarkably, both experimental datasets basically agree in the points farther from the stopping maximum
(at $v_\mathrm{H^+} \sim 1$~a.u. and $\sim 3.5$~a.u.),
but differ significantly around $v_\mathrm{H^+} \sim 2$~a.u.,
which is precisely the region where the electronic-mediated nuclear stopping found by us is relevant.
At this point, we stress that the electronic-mediated kinetic energy transfer from projectile to target nuclei
that we have just described occurs without any noticeable deflection of the projectile.
Regarding the experimental determination of the electronic stopping power
done via analyses of scattering spectra of the irradiating projectiles,
this implies that the stopping mechanism that we just described is
experimentally indistinguishable from the usually assumed
mechanism purely based on electronic excitations.
Therefore, we conclude that the `\textit{electronic} stopping power' experimentally reported
for H$_2$O and D$_2$O in that regime of projectile velocities in Refs.~\cite{wenzel_pr1952,bauer_nimphysresb1994}
is actually equivalent to the \textit{total} stopping power determined by us in this work.

Finally, the qualitative differences between our computational results and the experimental results deserve further comments.
The experimentally reported isotopic effects are greater than in our simulations, 
as clearly seen in Fig.~\ref{figtotelstop}. 
On one hand, this could be due to a trivial sampling issue:  $S_\mathrm{n}$ is much more
sensitive to small differences of the projectile trajectory than $S_\mathrm{e}$.
This suggests that our current estimation of the average $\langle S_\mathrm{n} \rangle$ may be improved 
with a more intensive sampling of the possible projectile trajectories,
maybe including directions not parallel to the hexagonal channel.
But on the other hand and more interestingly, the presence of H/D isotopic effects always open the question of the possible role 
of nuclear quantum effects such as the greater quantum delocalization of H compared to D. 
At that respect, we note that our simulations employ classical nuclei, and of course the inclusion 
of nuclear quantum effects in the simulations 
could
greatly affect the dynamical response of the 
H nuclei to the dramatic changes of the electronic density of the system during the irradiation.

\textit{Conclusions and outlook.--}~
Our extensive Ehrenfest MD simulations, which strictly consider the atomic forces during the irradiation process,
disclosed a so far unknown channel for the energy transfer from the projectile to the target nuclei. 
This transfer occurs due to the dramatic changes of the electronic density of the target, 
which are the origin of great forces on the target nuclei during the passing of the projectile.
This phenomenon is fundamentally different from other well-known processes such as the electron-phonon 
energy transfer that occurs after the irradiation at longer timescales.
The fact that this electronic-mediated nuclear stopping occurs at the short timescale of the irradiation itself
implies that only light atoms (i.e. hydrogen) experience a meaningful gain of the kinetic energy, 
and moreover is the reason of an isotopic effect when comparing the stopping of light vs.~heavy water, 
in agreement with experiments that had been so far unexplained. 
Indeed, despite the fact that this effect occurring with no deflection of the projectile makes it rather elusive when 
it comes to its experimental determination, the isotopic effect experimentally described for H$_2$O vs D$_2$O 
is a very strong confirmation of our simulation results.
Actually, the experimental data show an even stronger isotopic effect than our simulations, 
which could be due to 
the missing nuclear quantum effects 
in our approach.
This aspect opens up an interesting --though technically challenging-- avenue for further research.

We expect that this effect will be present in aqueous systems, and in particular biological tissues.
Regarding the latter, this could represent a further mechanism for the creation of damage in biomolecules,
which would add up to other purely electronic damage mechanisms already described~\cite{shepard_prl2023}.
On the other hand, it remains to be seen whether it will also occur in hydrogen-rich materials that lack the precise electronic structure of water
and its associated H-bond network (e.g. hydrocarbon-based polymers of industrial relevance).
Clearly, these findings open up a new and exciting aspect of the radiation damage processes that deserves to be explored.

\begin{acknowledgments}
This project has received funding from the European Union's Horizon 2020 research and innovation
programme under the Marie Sk\l{}odowska-Curie grant agreement No 748673.
This work received funding from the Research Executive Agency under the EU’s Horizon 2020 Research and
Innovation program ESC2RAD (Grant ID 776410).
We acknowledge PRACE for awarding us access to MareNostrum at Barcelona Supercomputing Center (BSC), Spain.
The author thankfully acknowledges the computer resources at MareNostrum and the technical support provided by 
Barcelona Supercomputing Center (RES-FI-2018-1-0024, RES-FI-2018-3-0031, RES-FI-2019-1-0037, RES-FI-2019-2-0017,
RES-FI-2019-3-0021, RES-FI-2020-3-0014, RES-FI-2021-1-0035,
RES-FI-2022-1-0028,
RES-FI-2022-3-0040).
This work has been supported by the Madrid Government (Comunidad de Madrid-Spain) 
under the Multiannual Agreement with Universidad Polit\'ecnica de Madrid 
in the line Support for R\&D projects for Beatriz Galindo researchers, 
in the context of the V PRICIT (Regional Programme of Research and Technological Innovation).
\end{acknowledgments}

\bibliography{references}

\end{document}